\documentstyle [twocolumn,epsf]{mn}
\newcommand{\mincir}{\raise
-2.truept\hbox{\rlap{\hbox{$\sim$}}\raise5.truept\hbox{$<$}\ }}
\newcommand{\magcir}{\raise
-2.truept\hbox{\rlap{\hbox{$\sim$}}\raise5.truept\hbox{$>$}\ }}
\newcommand{\minmag}{\raise
-2.truept\hbox{\rlap{\hbox{$<$}}\raise6.truept\hbox{$<$}\ }}
\newcommand{\be}{\begin{equation}}
\newcommand{\ee}{\end{equation}}
\newcommand{\ba}{\begin{eqnarray}}
\newcommand{\ea}{\end{esqnarray}}
\newcommand{\brr}{\begin{array}}
\newcommand{\err}{\end{array}}
\newcommand{\bc}{\begin{center}}
\newcommand{\ec}{\end{center}}

\renewcommand{\baselinestretch}{1.2}
\title[The PSCz Dipole Revisited]{The PSCz Dipole Revisited}
\author[S.Basilakos \& M.Plionis]
{Spyros Basilakos$^{1,2}$ \& Manolis Plionis$^{1,3}$ \\
$^1$Institute of Astronomy \& Astrophysics, National Observatory of
  Athens, I.Metaxa \& B.Pavlou, P.Penteli 152 36, Athens, Greece\\
$^{2}$ Academy of Athens, Research Center for Astronomy \& Applied
  Mathematics, Soranou Efessiou 4, 11-527, Athens, Greece\\
$^3$ Instituto Nacional de Astrof\'{\i}sica, \'Optica y Electronica (INAOE)
Apartado Postal 51 y 216, 72000, Puebla, Pue., M\'exico\\
}

\begin{document}

\maketitle

\begin{abstract}
We re-examine the gravitational acceleration (dipole) 
induced on the Local Group of galaxies by the Infrared Astronomical
Satellite (IRAS) galaxy distribution of
the Point Source Catalogue redshift survey (PSCz). 
We treat the cirrus-affected low galactic latitudes by
utilizing a spherical harmonic expansion of the galaxy surface 
density field up to the octapole order. 
We find strong indications for significant contributions to 
the Local Group motion from depths up to $\sim 185$ $h^{-1}$ Mpc and
possible contribution even from $\sim 210$ $h^{-1}$ Mpc,
in agreement with the recent analysis of Kocevski \& Ebeling of 
a whole sky X-ray cluster survey.
What changes with respect to the previous PSCz dipole analyses is: (a)
the large-scale dipole contributions and (b)
an increase of the overall dipole amplitude due 
to the important contribution of the local volume ($\mincir 4 \;h^{-1}$
Mpc), which we now take into account.
This results in a lower value of the
$\beta (\equiv \Omega_{\rm m}^{0.6}/b)$ parameter, which we find to be
$\beta_{\rm IRAS} \simeq 0.49$ in real space. Therefore, 
for the concordance cosmological model 
($\Omega_{\rm m}=1-\Omega_{\Lambda}=0.3$) 
the IRAS galaxies bias factor is $b_{\rm IRAS}\simeq 1$ which means that
IRAS galaxies are good traces of the underlying matter distribution. 

\vspace{0.25cm}

\end{abstract}

\section{Introduction}
It is well known that the peculiar velocity of the Local Group 
of galaxies is accurately determined from 
the CMB temperature dipole (Kogut et al. 1996; Bennett et al. 2003),
being $u_{LG}=$622 km/sec towards $(l,b)=(277^{\circ},30^{\circ})$.
Under the framework of linear theory, the most 
probable cause for this motion as 
well as for the observed peculiar motions of other galaxies and clusters  
is gravitational instability 
(see Peebles 1980). The latter is supported by the fact that the 
gravitational acceleration of the Local Group of galaxies, as traced by
many different samples of extra-galactic mass tracers is well aligned with
the general direction of the CMB dipole (eg. Yahil, Walker \& 
Rowan-Robinson 1986; Lahav 1987; Plionis 1988; 
Lynden-Bell et al. 1988; Miyaji \& Boldt 1990;
Rowan-Robinson et al. 1990; Strauss et al. 1992; Hudson 1993; 
Scaramella et al. 1991; Plionis \& Valdarnini 1991; Branchini \&
Plionis 1996; Basilakos \& Plionis 1998; Schmoldt et al. 1999;
Rowan-Robinson et al. 2000; Kocevski, Mullis \& Ebeling 2004;
Erdogd\v{u} et al. 2006; Kocevski \& Ebeling 2006). 

However, what has been debated over 
the last decades, is the largest depth from which density fluctuations 
contribute significantly to the gravitational field that shapes the Local Group
motion, which is called dipole convergence 
depth, $D_{\rm conv}$.

The outcome of many studies in the past, using 
different flux or magnitude limited galaxy and cluster samples, was  that 
the apparent value of the dipole convergence depth differed from 
sample to sample, in the range from 40 to 160 $h^{-1}$ Mpc, with a strong 
dependence to the sample's characteristic depth. 
For example, the dipole of the 
IRAS Point Source Catalogue for redshifts (hereafter
PSCz) was claimed to mostly converge within a depth of $\sim 100$
$h^{-1}$ Mpc with negligible, if any, contributions beyond $\sim 140$
$h^{-1}$ Mpc (Rowan-Robinson et al. 2000; Schmoldt et al. 1999), while that of
the optical Abell/ACO cluster sample, which is volume limited out to a large 
depth ($\magcir 200 \; h^{-1}$ Mpc), was claimed to converge
at $\magcir 160\; h^{-1}$ Mpc but with significant contributions beyond
$\sim 140$ $h^{-1}$ Mpc (Scaramella et al. 1991; Plionis \& Valdarnini 1991; 
Branchini \& Plionis 1996). The latter result has been 
confirmed using X-ray cluster samples, which are free of the various 
systematic effects from which the optical catalogues suffer (Plionis \& 
Kolokotronis 1998; Kocevski et al. 2004; Kocevski \& Ebeling 2006).
Of course, under the assumption that there is a linear bias 
relation between the cluster, the galaxy and the 
underlying matter density fluctuations, as proposed by Kaiser (1987), 
then the galaxy dipole should also show similarly deep contributions. 
This hinted that
the shallower galaxy dipole convergence was spurious, and it 
could be attributed to the lack of adequately sampling of the distant density
fluctuations. 

This discussion was recently re-opened by two important papers, that
of Erdogd\v{u} et al. (2006) who analysed the 2 Micron All-Sky Redshift
Survey (hereafter {\it 2MRS}) and
found $D_{\rm conv} \sim 60 \; h^{-1}$ Mpc, and
of Kocevski \& Ebeling (2006) who analysed a complete whole-sky
survey of X-ray clusters and found $D_{\rm conv} \magcir 200
\;h^{-1}$ Mpc.

Due to the fact that the Erdogd\v{u} et al. (2006) analysis accounted for
the dipole contribution of local galaxies with known true distances and
due to the potential importance of such contributions, especially on
the light of the apparent difference between the original PSCz 
and {\em 2MRS} dipole results (see section 5 in Erdogd\v{u} et
al. 2006), we
decided to re-analyse the PSCz catalogue along the same lines.
In particular, we estimate the PSCz dipole using 
an approach which is based on (a) a spherical harmonic expansion of 
the galaxy surface density  up to the octapole order in order to mask
the ``zone of avoidance''
and (b) accounting for contributions to the PSCz 
dipole from the nearby galaxies.


\section{Data and Method}
In this analysis we utilize the IRAS flux-limited 
60-$\mu$m redshift survey (PSCz) which is described in
Saunders et al. (2000). It is based on the IRAS Point Source 
Catalogue and contains $\sim 15000$ galaxies with flux $>0.6$ Jy. 
The subsample we use, defined by $|b|\ge 8^{\circ}$ and limiting galaxy 
distance of 240 $h^{-1}$ Mpc, contains $\sim 12300$ galaxies and 
covers $\sim 85\%$ of the sky.

We remind the reader that 
using linear perturbation theory one can relate the gravitational 
acceleration of an observer, induced by the
surrounding mass distribution, to her/his peculiar velocity, according
to:
\be
{\bf v(r)}=\frac{H_{0} \beta}{ 4 \pi \langle n \rangle} \int
\delta({\bf x})
\frac{\bf x}{|{\bf x}|^3} {\rm d}r = \beta {\bf D}(r) \;,
\ee
where $\beta=\Omega_{\rm m}^{0.6}/b$ and $\langle n \rangle$ is the 
mean number density. The dipole moment, ${\bf D}$, is estimated by weighting
the unit directional vector pointing to the position of each galaxy, with its
gravitational weight and summing over the galaxy distribution: 
\be
{\bf D} =\frac{H_{0}}{4 \pi \langle n \rangle} 
\sum \frac{\hat{{\bf r}}
}{\phi(r) \; r^{2}}  \;\;.
\ee
In order to estimate the dipole
using observational data, it is necessary to recover the true 
galaxy density field from the observed flux-limited samples. This is done
by weighting each galaxy by $\phi^{-1}(r)$, 
where $\phi(r)$ is the selection function, defined by:
\be
\phi(r)=\frac{1}{\langle n \rangle} \int_{L_{{\rm min}}(r)}^{L_{{\rm max}}} 
\Phi(L)\,{\rm d}L \;.
\ee
$\Phi(L)$ is the luminosity function of the objects under study,
$L_{\rm {min}}(r)=4\pi r^{2} S_{\rm {lim}}$, with $S_{\rm {lim}}$
the flux limit of the sample
and $\langle n \rangle$ is the mean tracer number density, given by 
integrating the luminosity function over the whole luminosity
range. Note, that here we use the 
Rowan-Robinson el al. (2000) luminosity function. 

Finally, due to discreteness 
effects and the steep selection function with depth, there is
an additive dipole term, the shot-noise dipole, for which we 
have to correct our 
raw dipole estimates. In this work, we estimate the shot noise dipole using
the analytic formula of Strauss et al. (1992) and Hudson (1993). 
Assuming Gaussianity, the Cartesian 
components of the shot noise dipole are equal 
and thus $\sigma_{\rm sn}^{2}= 3 \sigma_{i,1D}^{2}$. 
Choosing the coordinate system such that one of the shot-noise dipole components
is parallel to the z-axis of the true dipole we use an approximate
correction model of the raw dipole according to (see
Basilakos \& Plionis 1998):  
\be
D_{\rm cor}=D_{\rm raw}-\sigma_{\rm sn}/\sqrt{3}\;.
\ee

\noindent

\subsection{Treatment of the Masked regions}
Firstly, we need to model the excluded, due to cirrus emission,
galactic plane and correct accordingly our raw dipole.
We do so by extrapolating to these regions the galaxy distribution from the rest of the unit 
sphere with the help of a spherical harmonic expansion of the galaxy surface 
density field (e.g., Yahil, Walker, Rowan-Robinson 1986; Lahav 1987; 
Plionis 1988; Tadros et al. 1999). For the purpose of the present 
analysis we expand the PSCz surface density field,
$\Sigma(\theta,\phi)$, in spherical harmonics up to the 
octapole order:
\begin{eqnarray*}
\Sigma(\theta,\phi)=\sum_{l=0}^{3}A_{l}^{0}P_{l}(x)+ 
\sum_{m=1}^{l} P_{l}^{m}(x) [ A_{l}^{m}{\rm cos}(m\phi)+ \\
B_{l}^{m}{\rm sin}(m\phi) ]
\end{eqnarray*}
where $0\le \theta \le \pi$ ($\theta=90^{\circ}-b$), 
$0\le \phi \le 2\pi$ and $P_{l}^{m}(x)$ 
the associated Legendre functions ($x={\rm cos} \theta$).
The observed surface density is related to 
the intrinsic one, $\sigma(\theta,\phi)$, by:
$\Sigma(\theta,\phi)=M(\theta)\sigma(\theta,\phi)$, 
where $M(\theta)$ is a mask to account for the excluded
galactic plane:
$$
M(\theta)=\left\{ \begin{array}{cc}
                     1 & \mbox{for $\mid b \mid \ge b_{\rm lim}$} \\
		     0 & \mbox{for $\mid b \mid < b_{\rm lim}$}
		     \end{array}
                      \right.  
$$ 
with $b_{\rm lim}=8^{\circ}$.
Thus the overall problem is reduced to the inversion of a 
$16\times 16$ matrix, ${\bf T}$, which then provides the values
for the whole sky components of the galaxy distribution:
$\bf { C=T^{-1} A}$, 
where ${\bf A}$ is a $16\times 1$ matrix which contains 
the observed components $(A_{l}^{m},B_{l}^{m})$
from the incomplete data and ${\bf C}$ is a $1\times 16$ 
matrix containing the model corrected $(\alpha_{l}^{m},b_{l}^{m})$
components. 
Then the differential dipole components, 
at each spherical shell in $r$, are corrected according to this 
expansion.

Secondly, about $\sim 4\%$ of 
the sky was unobserved  by IRAS and we therefore apply 
to these areas a homogeneous distribution of galaxies following the
PSCz redshift selection function.

\subsection{Determining distances from redshifts}
All heliocentric redshifts are first transformed to the Local Group frame
using $c z \simeq c z_{\odot}+300 \sin(l)\cos(b)$. 
We then derive the distance of each tracer by using:
$$
r=\frac{2 c}{H_{\circ}} \left[1-(1+z-\delta z)^{-1/2} \right]
(1+z-\delta z)^{3/2} \;,
$$
with $H_{0}=100 \;h$ km/sec Mpc and 
$$\delta z =\frac{1}{c} [{\bf u}(r)-{\bf u}(0)] \cdot \hat{r}$$
the non-linear term that takes
into account the contribution of galaxy peculiar velocities, 
${\bf u}(r)$, to the galaxy redshift [${\bf u}(0)$ the peculiar velocity of
the Local Group].
Instead of using 3D reconstruction schemes (e.g., Schmoldt et al. 1999;
Branchini \& Plionis 1996; Branchini et al. 1999; Rowan-Robinson et al. 2000) 
to estimate this term, we use a rather crude but adequate for the
purpose of this work model velocity field (see Basilakos \& Plionis
1998), the main assumptions of which are:

\noindent
{\bf (a)} The tracer peculiar velocities can be split 
in two vectorial components; that of a bulk flow and of a local non-linear term:
${\bf u}(r)={\bf V}_{\rm bulk}(r)+{\bf u}_{\rm nl}(r)$

\noindent
{\bf (b)} The first component dominates and thus
${\bf u}(r) \cdot \hat{\bf r} \approx {\bf V}_{\rm bulk}(r) \cdot \hat{\bf r}$.
As for the observed bulk flow direction and profile
we use that given by Dekel (1997) and combined with 
that of Branchini, Plionis \& Sciama (1996).
The zero-point, $V_{\rm bulk}(0)$, and the direction of the bulk flow is
estimated assuming, due to the ``coldness'' 
of the local velocity field (eg. Peebles 1988), by
${\bf V}_{\rm bulk}(0)={\bf u}(0) - {\bf u}_{\rm inf}$ 
(where $u_{\rm inf}= 200$ km/s is the LG in-fall velocity to 
the Virgo Cluster).

\subsection{Treatment of the Local Volume}
As in Erdogd\v{u} et al. (2006) we use the true distances for
all possible nearby PSCz galaxy. To this end we  
cross correlate the PSCz catalogue ($|b|\ge 8^{\circ}$)
with two literature datasets of neighbouring galaxies with measured 
distances (Freedman et al. 2001; Karachentsev et al. 2004), which revealed 
56 common galaxies up to $cz \le 1000$ km/s, 
out of which three are blue-shifted galaxies. 
Two of these
belong to the Virgo cluster area and thus we put them to 
the center of Virgo 
($15.4$ Mpc), while the other one is assigned to the zero velocity 
surface $\sim 1.18 \;h^{-1}$Mpc (see Courteau \& van den Bergh 1999;
Erdogd\v{u} et al. 2006). 

\begin{figure}
\mbox{\epsfxsize=11cm \epsffile{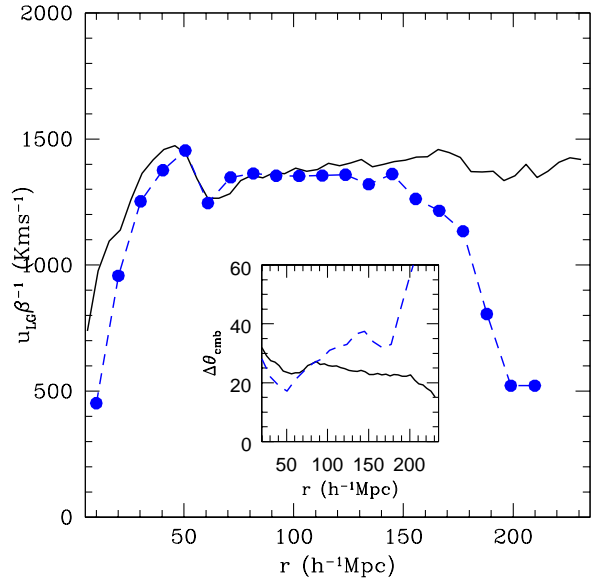}}
\caption{Comparison of our PSCz (solid line) and the 
{\it 2MRS} (dashed line) number-weighted dipoles in redshift space,
corrected for their respective shot-noise according to eq. (4).
The insert panel shows 
the misalignment angles (ours -solid line and 
Erdogd\v{u} et al. 2006 - dashed line) between the galaxy and 
CMB dipoles as a function of distance from the
Local Group.} 
\end{figure}

\section{Results}
\subsection{Redshift-Space}
In figure 1 we present a comparison in redshift space
of our PSCz (solid line) 
and the {\it 2MRS} (dashed line; Erdogd\v{u} et al. 2006)
number-weighted galaxy dipoles, 
out to 210 $h^{-1}$ Mpc, both corrected for their respective
shot-noise (eq. 4).
The corresponding galaxy-CMB dipole misalignment angle
is shown in the insert of figure 1. 
The shape and amplitude of the two profiles are in very good agreement 
but only within $\mincir 100-120 \; h^{-1}$ Mpc.
The {\it 2MRS} dipole seemingly reaches a
plateau at 60$h^{-1}$Mpc while the corresponding PSCz
dipole continues to grow.

Beyond $\sim 140 \; h^{-1}$ Mpc, however, the {\it 2MRS} dipole drops 
dramatically, a behaviour not seen in the
flux-weighted {\em 2MRS} dipole (see Erdogd\v{u} et al. 2006),
while the PSCz dipole appears to increase by $5\%$ up to
$\sim 170\;h^{-1}$Mpc.  Despite, the fact that the 
{\it 2MRS} catalogue samples the volume
within $\sim 140h^{-1}$Mpc better than the PSCz survey, the opposite
is observed at greater distances,
as revealed by comparing the two redshift
distributions (see figure 3 of Erdogd\v{u} et al. 2006).
Therefore, the PSCz catalogue samples the
distant matter fluctuations better than the {\it 2MRS} survey. 
This can also be appreciated when comparing the corresponding 
PSCz and {\em 2MRS} dipole misalignment angles with the CMB  
(see the insert panel of figure 1).

\subsection{Real-Space}
We concentrate now on the real space dipoles. Due to the fact that 
Erdogd\v{u} et al. (2006) did not present the {\it 2MRS} dipole 
in real space, we will compare our results only with previous PSCz analyses
(Schmoldt et al. 1999; and Rowan-Robinson et al. 2000).
These studies, however, utilize different techniques 
in order to correct for the redshift space
distortions while they also practically exclude the local volume 
(galaxies with $r\le 4h^{-1}$Mpc), an important 
contributor of the local velocity field. 

\begin{figure}
\mbox{\epsfxsize=11cm \epsffile{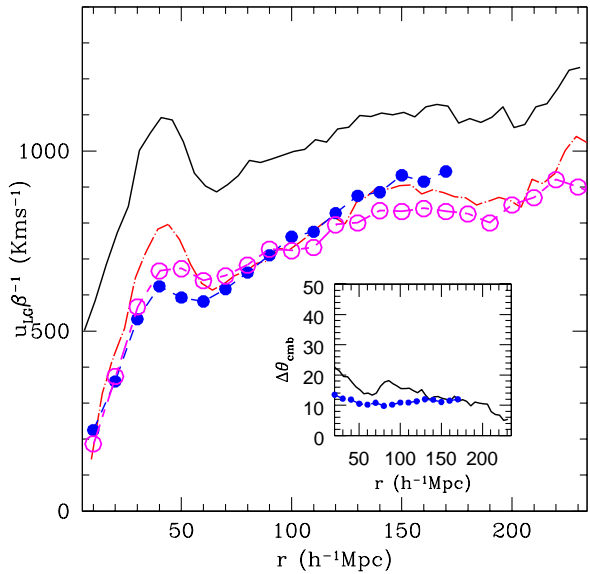}}
\caption{The cumulative PSCz dipole in real space (a) taking
  into account contributions from the local volume (solid line)
and (b) excluding the local volume, $r\le 4\;h^{-1}$Mpc (dot dashed
  line). The open and the solid points show the 
Rowan-Robinson et al. (2000) and Schmoldt et al. (1999)  
dipole, respectively. The insert panel shows 
the misalignment angles (ours -solid line and 
Schmoldt et al. 1999 - solid points) between the PSCz and 
CMB dipoles as a function of distance from the
Local Group.}
\end{figure}
In figure 2 we present
our PSCz dipole (solid line) in real space taking into account the
contributions from the local volume.  
Apparently, there are three amplitude dipole bumps. 
The first one, at $\sim 40 \;h^{-1}$ Mpc, is caused by the Great
Attractor (Lynden-Bell et al. 1988). The second one is caused by
the Shapley concentration 
(Scaramella et al. 1989; Raychaudhury 1989; Bardelli et al. 1994), 
a huge mass overdensity 
located beyond $\sim 130$ $h^{-1}$ Mpc in the general direction of the 
Hydra-Centaurus supercluster
(e.g., Plionis \& Valdarnini 1991; Scaramella et al. 1991; 
Branchini \& Plionis 1996; Plionis \& Kolokotronis 1998; 
Kocevski et al. 2004; 
Kocevski \& Ebeling 2006 and references therein). 
Finally, from our analysis there appears to be also a third bump,
at $r\magcir 210h^{-1}$Mpc. However, due to sparse
sampling such deep contributions cannot 
be accurately determined by the present galaxy sample (e.g.,
Kolokotronis et al. 1996).

If we now exclude the local volume ($r\mincir 4 \;h^{-1}$ Mpc), we can directly
compare our estimated PSCz dipole (dot dashed line) with those derived 
by Schmoldt et al. (1999) and 
Rowan-Robinson et al. (2000) (solid and open circles respectively). 
The three dipoles compare very well, although they are based on a
different treatment of the masked regions and of the method used to
translate redshift to real space, to which we should attribute the
small differences in the Local Universe ($r\le
60 \;h^{-1}$Mpc) and between 140 and 200 $h^{-1}$ Mpc (in which
distance range we slightly differ only with Rowan-Robinson et
al. results).
The insert panel of figure 2 shows
the PSCz-CMB dipole misalignment angle as a function of
distance from the Local group for our (line) and Schmoldt et
al. (dots) results. Evidently, they almost coincide at large
distances. The further decrease of our PSCz-CMB dipole misalignment angle out to
$\sim$240 $h^{-1}$ Mpc supports the existence
of possible dipole contributions from very large depths.

\subsection{Robustness of Spherical Harmonic Mask}
As we discussed in section 2.1 we have decided to use the
mathematically elegant spherical harmonic apporach in order to deal
with the so-called ``zone of avoidance''. For concistency 
we have decided to use the same
galactic latitude limit ($b_{\rm lim}=8^{\circ}$) as in Schmoldt et
al. (1999), who however used a cloning technique (see Lynden-Bell,
Lahav \& Burstein 1989) to extrapolate the
galaxy distribution of two strips, above and below the ``zone of avoidance'',
to the masked region. Rowan-Robinson et al. (2000) used a slightly
different technique although of the same philosophy. They divided the
sky into 413 areas, each approximately 100 deg$^{2}$. The areas
affected by the mask where artifiscially filled, at random positions, 
with flux-velocity
pairs either randomly selected from the whole data set or from two
neighboring bins that were at least 75\% full. 
However, the resulting dipole direction and amplitude was found, especially at large
distances, to be distinctly different.

\begin{figure}
\mbox{\epsfxsize=11cm \epsffile{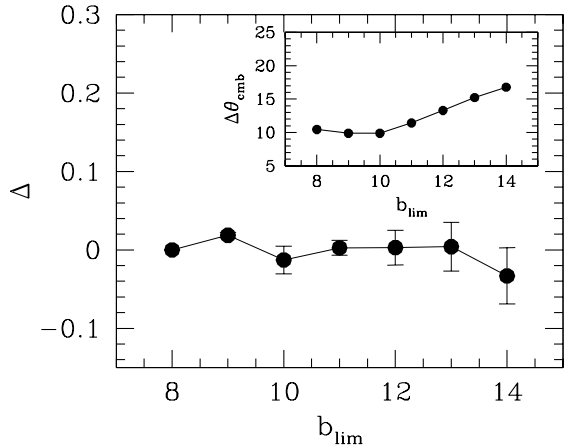}}
\caption{Fluctuations of the real-space PSCz 
dipole amplitude around its value at
  $r=200 \; h^{-1}$ Mpc as a function of the limiting value of the
  galactic latitude defining the masked region. {\sc Insert Panel:} The
  corresponding dipole misalignment angles as a function of $b_{\rm lim}$.}
\end{figure}

Therefore it appears imperative
to test the robustness of our approach, to variations of the
galactic latitude limit. In Fig. 3 we present the fluctuation of the
real-space dipole amplitude, $\Delta$, around its value at $r=200 \; h^{-1}$ Mpc for the
different $b_{\rm lim}$ values. The robusteness of the dipole amplitude
is evident (variations $< 3\%$). As for the dipole misalignment angle
with respect to the CMB (see insert of Fig. 3) it is always lower than
$\sim 16^{\circ}$. However, there is a small increase of the misalignment angle as
a function of increasing $b_{\rm lim}$ which should be expected from the 
important structures located
near the galactic plane (eg. Radburn-Smith et al. 2006 and references threin).

We conclude that our dipole results are robust to small variations of
the considered size of the ``zone-of -avoidance''.

\subsection{Deep Contributions}
In order to further investigate the probable deeper PSCz dipole contributions we
estimate the differential dipole in equal volume shells of
different sizes (see 
Plionis, Coles \& Catelan 1993; Basilakos \& Plionis 1998). 
In table 1, we present results for the case of $\delta V \simeq 1.4 \times
10^{5} \; h^{-3}$ Mpc$^{3}$.
A measure of the significance of the differential dipole 
of the individual shells, is given by:
\be 
\frac{S}{N}=\frac{D_{\rm raw}}{\sigma_{\rm 3D}} \cos(\delta\theta_{\rm
cmb}) \;.
\ee
We observe that two local shells ($< 129\; h^{-1}$ Mpc) and 
four distance shells reaching up to 228 $h^{-1}$ Mpc have 
 signal to noise ratios $>5$ and relatively
 small misalignment angles between the differential and CMB dipoles:
$\delta\theta_{\rm cmb} \mincir 24^{\circ}$ 
and $24^{\circ} \mincir \delta\theta_{\rm cmb} \mincir
56^{\circ}$ for the inner and outer shells, respectively.
The joint probability to have random alignments within the above 
misalignment limits is extremely low. Indeed,
the formal probability that two vectors are 
aligned within $\delta\theta_{\rm cmb}$
is given by the ratio of the solid angle which 
corresponds to $\delta\theta_{\rm cmb}$, to 
the solid angle of the whole sphere, ie.,
$p_{f}(\delta\theta_{\rm cmb}) = \sin^{2}(\delta\theta_{\rm cmb}/2)$.

We can now estimate the random joint probability of alignment, 
within the observed $\delta\theta_{\rm cmb}$, of $N$ independent vectors, 
which is given by:
$P_{N} \approx \prod_{i=1}^{N} p_{i}(\delta\theta)/p_{i}(90^{\circ})$.
Due to (a) the fact that galaxies are correlated spatially and 
(b) the vicinity of some of the shells,
we consider the joint probability of the first (1-103 $h^{-1}$ Mpc), the eighth (187-196
$h^{-1}$ Mpc) and the tenth (214-221 $h^{-1}$ Mpc) shells, for which we
find $P_{3} \mincir 5 \times 10^{-4}$. This should be considered
as a conservative upper limit since we have not taken into account the
other four, $S/N>5$, shells.
\begin{table}
\caption[]{Differential PSCz dipole directions, 
misalignment angles with respect to the CMB dipole, differential dipole signal to 
noise ratio and probabilities of alignment within 
$\delta\theta_{\rm cmb}$ (see text for complete definition).}

\tabcolsep 5pt
\begin{tabular}{ccccccc} \hline
$h^{-1}$Mpc &  $N_{\rm gal}$ & $S/N$ & $l^{\circ}$ & $b^{\circ}$  & 
$\delta\theta_{\rm cmb}$ & $p_f$ \\ \hline
      1-103&  8238 & 141.7 &  290.7 &    40.3&   15&    0.017\\
    103-129&  1390&   13.1&   249.8&     39.3&   24&    0.043\\
    129-148&   662&    1.8&   191.4&     15.7&   78&    0.400\\
    148-163&   480&    3.1&   324.9 &   -14.2&   64&    0.280\\
    163-176&   375&   -3.8&   147.7  &  -34.9&  137&    0.868\\
    176-187&   260&    0.2&   304.9  &  -54.4&   88&    0.480\\
    187-196&   201&    5.2&   300.6  &   44.8&   24&    0.042\\
    196-205&   164&   -5.9&   132.8&    -25.8&  148&    0.925\\
    205-214&   165&    5.1&   254.8 &   -21.5&   56&    0.219\\
    214-221&   142&    9.3&   297.3  &    1.5&   34&    0.088\\
    221-228&   107&    9.8&   269.8 &   -10.5&   41&    0.123\\
    228-235&   100&    1.6&   325.6 &   -36.1&   80&    0.416\\ \hline
\end{tabular}
\end{table}
As a further test we also use a Monte-Carlo procedure 
(see Basilakos \& Plionis 1998) 
to test whether the dipole-CMB alignments could be induced 
due to our frame transformation procedure. We find that 
the probability to have random alignments due to 
the frame transformation is low and comparable to $p_{f}$.
As a result, we conclude that the differential 
dipole directions are not randomly oriented
with respect to the CMB and therefore we have further indications 
for significant dipole contributions from large depths.

The deeper differential dipole signal supports 
the existence of dipole contributions from large depths, were
numerous groupings and cluster of galaxies have been found (Kocevski \& Ebeling 2006).
This result is however in contrast with the conclusions of Rowan-Robisnon et
al. (2000) and Schmoldt et al. (1999), who find negligible, if any,
dipole contributions beyond $\sim 140$ $h^{-1}$ Mpc. Our different
conclusion is based, we beleive, in the different approach
(differential dipole) used to
investigate the possible deep dipole contributions in the presence of
sparse data.

\subsection{The $\beta$ Parameter}
Under the biasing ansatz it is easy to
obtain the value of the $\beta(\equiv \Omega_{\rm m}^{0.6}/b)$ parameter.
Erdogd\v{u} et al. (2006) found 
$\beta_{\rm 2MRS}\simeq
0.40$ from the ${\it 2MRS}$ dipole in redshift space. However, using
the corrected for shot-noise ${\it 2MRS}$ dipole (according to eq. 4) one
finds: $\beta_{\rm 2MRS}\simeq 0.44$, which is exactly the value that
we find also from our PSCz analysis.
In real space our PSCz dipole gives 
$\beta_{\rm IRAS}\simeq 0.49$, which means that 
in the framework of the concordance 
cosmological model ($\Omega_{\rm m}=1-\Omega_{\Lambda}=0.3$) 
the IRAS galaxy bias factor is $b_{\rm IRAS}\simeq 1$. It is interesting to
mention that our results are in agreement with those derived  
from the so called VELMOD technique using a variety of extragalactic data sets 
(see Davis, Nusser \& Willick 1996; Willick \& Strauss 1998; 
Nusser et al. 2001; Pike \& Hudson 2006). Also our $\beta$ 
results are in agreement with those found by 
Maller et al. (2003), based on the {\it 2MASS} flux
weighted dipole.

If we take out the contributions from the local volume the 
$\beta_{\rm IRAS}$ parameter is found to be $\simeq 0.7$ 
in agreement with Basilakos \& Plionis (1998), 
Schmoldt et al. (1999), Rowan - Robinson et al. (2000) and 
Ciecielag \& Chodorowski (2005).

\section{Conclusions}
We have revisited the PSCz dipole using a spherical harmonics 
expansion of the galaxy density field up to the octapole order in
order to
model the excluded low-galatic latitudes.
We have also used measured distances 
from the literature for the nearby galaxies and we find that 
the amplitude of the dipole increases with respect to the 
previous PSCz dipole analysis.
We also find indications for 
significant contributions to the gravitational field that shapes
the Local Group motion from very large distances 
in agreement with a recent analysis of a deep all-sky X-ray
cluster survey. Finally, within the linear biasing ansatz we find 
$\beta_{\rm IRAS} \simeq 0.44$ and 0.49 in redshift and real 
space, respectively. This implies that 
within the framework of the concordance 
cosmological model ($\Omega_{\rm m}=1-\Omega_{\Lambda}=0.3$) 
the linear biasing factor of the IRAS galaxies is
$b_{\rm IRAS}\simeq 1$.



{\small 
\begin{thebibliography}{}
\bibitem[]{} Bardelli, S., Zucca, E., Vettolani, G., Zamorani, G.,
Scaramella, R., Collins, C.A., MacGillivray, H.T., 1994, MNRAS, 267, 665
\bibitem[]{}Basilakos, S. \& Plionis, M., 1998, MNRAS, 299, 637
\bibitem[]{}Bennett, C.L., et al., 2003, ApJS, 148, 1
\bibitem[]{}Branchini E., \&, Plionis M., 1996, ApJ, 460, 569
\bibitem[]{}Branchini E., Plionis M., \&, Sciama D.W., 1996, ApJ, 461, L17
\bibitem[]{}Branchini E., et al., 1999, MNRAS, 308, 1
\bibitem[]{}Ciecielag, P., \&, Chodorowski, M. J., 2004, MNRAS, 349, 945
\bibitem[]{}Courteau, S., \&, van den Bergh, S., 1999, AJ, 118, 337
\bibitem[]{}Davis, M., Nusser, A., \&, Willick, J. A., 1996, ApJ, 473, 22
\bibitem[]{}Dekel A., 1997, in L.da Costa ed. `Galaxy Scaling Relations: 
Origins, Evolution \& Applications', Springer, p.245 
\bibitem[]{}Erdogd\v{u}, P, et al., 2006, MNRAS, 368, 1515
\bibitem[]{}Freedman, W. L., et al., 2001, ApJ, 553, 47
\bibitem[]{}Hudson M.J., 1993, MNRAS, 265, 72
\bibitem[]{}Kaiser N., 1987, MNRAS, 227, 1
\bibitem[]{}Karachentsev, I. D., Karachentseva, V. E., Huchtmeier,
  W. K., \&, Makarov, D. I., 2004, AJ, 127, 2031
\bibitem[]{} Kocevski, D. D., Mullis, C. R.., \&, Ebeling, H., 2004, ApJ,
  608, 721
\bibitem[]{} Kocevski, D. D., \& , Ebeling, H., 2006, ApJ, 645, 1043
\bibitem[]{} Kogut et al., 1993, ApJ, 419, 1
\bibitem[]{}Kolokotronis, V., Plionis, M., Coles, P., Borgani, S., 
\&, Moscardini, L., 1996, MNRAS, 280, 186
\bibitem[]{}Lahav O., 1987,  MNRAS, 225, 213
\bibitem[]{}Lynden-Bell, D., Faber, S.M., Burstein, D., Davies, R.L., 
Dressler, A., Terlevich, R.J. \& Wegner, G. 1988, ApJ, 326, 19
\bibitem[]{} Lynden-Bell, D., Lahav, O. \& Burstein, D. 1989, MNRAS,
  241, 325
\bibitem[]{}Maller, A. H., McIntosh, D. H., Katz, N., \&,  
Weinberg, M. D., 2003, ApJ, 598, L1
\bibitem[]{}Miyaji T., \&, Boldt E., 1990, ApJ, 353, L3
\bibitem[]{}Nusser, A., et al., 2001, MNRAS, 320, L21
\bibitem[]{}Peebles P.J.E., 1980. The Large Scale Structure of the
Universe, Princeton University Press, Princeton New Jersey 
\bibitem[]{} Peebles, P.J.E., 1988, ApJ, 332, 17
\bibitem[]{} Pike R. W., \&, Hudson M. J., 2006, ApJ, in press, 
astro-ph/0511012
\bibitem[]{}Plionis, M., 1988, MNRAS, 234, 401 
\bibitem[]{}Plionis, M. \& Valdarnini, R., 1991, MNRAS, 249, 46
\bibitem[]{}Plionis M., Coles P., \&, Catelan P., 1993,  MNRAS, 262, 465
\bibitem[]{}Plionis M. \& Kolokotronis, V., 1998, ApJ, 500, 1
\bibitem[]{}Radburn-Smith, D.J., Lucey, J.R., Woudt, P.A.,
  Kraan-Korteweg, R.C., Watson, F.G., 2006, MNRAS, 369, 1131 
\bibitem[]{}Raychaudhury, S., 1989, Nat, 342, 251
\bibitem[]{}Rowan-Robinson M., et al., 1990,  MNRAS, 247, 1
\bibitem[]{}Rowan-Robinson M. et al., 2000, MNRAS, 314, 375
\bibitem[]{}Saunders W., et al. 2000, MNRAS, 317, 55 
\bibitem[]{}Scaramella, R., Baiesi-Pillastrini, G., Chincarini, G.,
Vettolani, G. Zamorani, G., 1989, Nat, 338, 562
\bibitem[]{}Scaramella, R., Vettolani, G., \&, Zamorani, G., 1991, ApJ, 376, L1
\bibitem[]{}Schmoldt, I. M., et al., 1999, MNRAS, 304, 893
\bibitem[]{}Strauss M.A., Yahil A., Davis M., Huchra J.P., \&, Fisher K., 1992
ApJ, 397, 395
\bibitem[]{}Tadros, H., et al., 1999, MNRAS, 305, 527
\bibitem[]{}Willick, J. A., \&, Strauss, M. A., 1998, ApJ, 507, 64
\bibitem[]{}Yahil A., Walker, D., \&, Rowan-Robinson, M., 1986, ApJ, 301, L1

\end{thebibliography}
}
\end{document}